\documentstyle[amssymb,aps,twocolumn]{revtex}

\draft

\begin{document}
\title{Low energy exciton states in a nanoscopic semiconducting ring}
\author{Hui Hu$^a$, Guang-Ming Zhang$^{a,b}$, Jia-Lin Zhu$^{a,b}$, Jia-Jiong Xiong$%
^a $}
\address{$^a$Department of Physics, Tsinghua University, Beijing 100084, P. R. China\\
$^b$Center for Advanced Study, Tsinghua University, Beijing 100084, P. R.
China}
\date{\today}
\maketitle

\begin{abstract}
We consider an effective mass model for an electron-hole pair in a
simplified confinement potential, which is applicable to both a nanoscopic
self-assembled semiconducting InAs ring and a quantum dot. The linear
optical susceptibility, proportional to the absorption intensity of
near-infrared transmission, is calculated as a function of the ring radius $%
R_0$. Compared with the properties of the quantum dot corresponding to the
model with a very small radius $R_0$, our results are in qualitative
agreement with the recent experimental measurements by Pettersson {\it et al}%
. \newline
{PACS numberes: 73.20.Dx, 71.35.-y, 78.66.Fd} \newline
\end{abstract}



Recent progress in nanoscopic fabrication techniques has made it possible to
construct self-assembled semiconducting nano-rings inside a completed
field-effect transistor \cite{lorke98,lorke99,lorke00,warburton}. The
nanoscopic rings may be considered as the best candidate to display various
pure quantum effects, as the nano-rings are in the scattering free and few
particles limit. By using two complementary spectroscopic techniques, Lorke
and collaborators performed the first spectroscopic measurements on
semiconducting InGaAs nano-rings occupied one or two electrons \cite
{lorke98,lorke99,lorke00}, and the experimental results have attracted a lot
of theoretical interests at the moment \cite
{chak,gudm,wendler,halonen,emperador,hu}. However, the spectroscopic data of
exciton effects in InGaAs nano-rings have also been obtained \cite
{pettersson} and clearly exhibited different excitonic properties from the
corresponding quantum dots \cite{magn,que,halonen92,wojs,hawprb99}. For
these experimental results, so far there has been no theoretical analysis
yet.

In this paper, we attempt to describe qualitatively the low-lying states and
some spectroscopic properties of the semiconducting nano-ring of InAs. Based
on an effective mass model Hamiltonian for an electron-hole pair in a
simplified confinement potential, the model Hamiltonian is separated into
the motion of center of mass, relative motion of the electron-hole pair, and
the mixed part. By varying radius of the ring $R_0$, both the quantum dots
and nano-rings are investigated within the same framework of a theory. In
the subspace of the zero total angular momentum $L=0$, the energy spectrum
of the exciton is derived, and the linear optical susceptibilities for the
excitons with a heavy hole or a light hole are calculated, respectively.
There exist significant differences of the excitonic properties between the
nano-rings and quantum dots.

The nanoscopic semiconducting ring is described by an electron-hole pair ($%
i=e,h$) with an effective band edge mass $m_i^{*}$ moving in a x-y plane.
The ring-like structure is well described by a double well potential, $%
U\left( {\bf \vec{r}}_i\right) =\frac 1{2R_0^2}m_i^{*}\omega _i^2\left( {\bf 
\vec{r}}_i{}^2-R_0^2\right) ^2$, which reproduces a soft barrier $\frac{%
m_i^{*}\omega _i^2R_0^2}2$\ at the center of the sample produced by
self-assembly \cite{lorke00,warburton}. Here, $R_0$\ is the radius of the
ring, $\omega _i$\ is the characteristic frequency of the radial
confinement, and a ring width{\bf \ }$W\approx 2\sqrt{\frac \hbar {%
2m_i^{*}\omega _i}}$. The model Hamiltonian is thus given by: 
\begin{equation}
{\cal H}=\sum\limits_{i=e,h}\left[ \frac{{\bf \vec{p}}_i^2}{2m_i^{*}}%
+U\left( {\bf \vec{r}}_i\right) \right] -\frac{e^2}{4\pi \varepsilon
_0\varepsilon _r\left| {\bf \vec{r}}_e-{\bf \vec{r}}_h\right| },
\end{equation}
where ${\bf \vec{r}}_i=\left( x_i,y_i\right) $ and ${\bf \vec{p}}_i=-i\hbar 
{\bf \vec{\nabla}}_i$ denote the position vector and momentum operator, $%
\varepsilon _0$ is the vacuum permittivity, and $\varepsilon _r$ is the
static dielectric constant of the host semiconductor. It should be pointed
out that the present {\it double-well} like confinement potential can be
rewritten as $U\left( {\bf \vec{r}}_i\right) =\frac 12m_i^{*}\omega
_i^2\left( r_i{}-R_0\right) ^2\frac{\left( r_i{}+R_0\right) ^2}{R_0^2}$. If
one replaces the operator $r_i$ in factor $\frac{\left( r_i{}+R_0\right) ^2}{%
R_0^2}$ by its mean value $\left\langle r_i\right\rangle =R_0$, the
confinement potential returns to the widely used parabolic form \cite
{lorke00,emperador,hu}. On the other hand{\bf ,} in the limit of the small
radius $R_0$ or for a small potential strength $\omega _i$, the soft barrier
at the ring center is very weak, and the description of the nano-ring is
more close to that of a quantum dot (see Fig. 1). For a fixed ring width (or
potential strength), the crossover from nano-rings to quantum dots is
determined by $R_0\sim \frac{\sqrt{2}}2W$\ or $\frac{\hbar \omega _i}2\sim 
\frac{m_i^{*}\omega _i^2R_0^2}2$, which means the lowest energy of radial
confinement is comparable to the soft barrier at the ring center{\bf .} It
should also be pointed out that our double-well confinement potential has
been used to calculate the far-infrared spectroscopy for a two-electron
nano-ring \cite{unpublish}, in good agreement with the recent experiment
done by Lorke{\bf \ }{\em et al}{\bf . }\cite{lorke00}.

Introducing the relative coordinate ${\bf \vec{r}}={\bf \vec{r}}_e-{\bf \vec{%
r}}_h$ and center-of-mass coordinate ${\bf \vec{R}}=\frac{m_e^{*}{\bf \vec{r}%
}_e+m_h^{*}{\bf \vec{r}}_h}{m_e^{*}+m_h^{*}}$, the model Hamiltonian is
divided into 
\begin{eqnarray}
H &=&{\cal H}_{cm}\left( {\bf \vec{R}}\right) +{\cal H}_{rel}\left( {\bf 
\vec{r}}\right) +{\cal H}_{mix}\left( {\bf \vec{R},\vec{r}}\right) , 
\nonumber \\
H_{cm} &=&\frac{{\bf \vec{P}}_{cm}^2}{2M}+\frac{M\omega _{cm}^2}{2R_0^2}%
\left( {\bf \vec{R}}{}^2-R_0^2\right) ^2,  \nonumber \\
H_{rel} &=&\frac{{\bf \vec{p}}_{rel}^2}{2\mu }+\frac \mu 2\frac{\left(
m_h^{*3}\omega _e^2+m_e^{*3}\omega _h^2\right) }{M^3R_0^2}r^4-\mu \omega
_{rel}^2{}r^2  \nonumber \\
&&-\frac{e^2}{4\pi \varepsilon _0\varepsilon _rr},  \nonumber \\
H_{mix} &=&-2\mu \left( \omega _e^2-\omega _h^2\right) \left( {\bf \vec{R}%
\cdot \vec{r}-}\frac{{\bf \vec{R}}^3{\bf \cdot \vec{r}}}{R_0^2}\right) 
\nonumber \\
&&+\frac{\mu \omega _{rel}^2}{R_0^2}\left[ R^2r^2+2\left( {\bf \vec{R}\cdot 
\vec{r}}\right) ^2\right]  \nonumber \\
&&+2\mu \frac{\left( m_h^{*2}\omega _e^2-m_e^{*2}\omega _h^2\right) }{%
M^2R_0^2}{\bf \vec{R}\cdot \vec{r}}^3,
\end{eqnarray}
where $\mu =\frac{m_e^{*}m_h^{*}}M$ is the electron-hole reduced mass and $%
M=m_e^{*}+m_h^{*}$ is the total mass. We have also introduced a
center-of-mass frequency $\omega _{cm}=\sqrt{\frac{\left( m_e^{*}\omega
_e^2+m_h^{*}\omega _h^2\right) }M}$ and a relative frequency $\omega _{rel}=%
\sqrt{\frac{m_h^{*}\omega _e^2+m_e^{*}\omega _h^2}M}$. The main advantage of
the separation of center-of-mass and relative coordinates is that the
negative Coulomb interaction $-\frac{e^2}{4\pi \varepsilon _0\varepsilon _rr}
$\ appears in{\bf \ }$H_{rel}$ only{\bf , }and the well-known
poor-convergence of the parabolic basis is thus avoided when the
characteristic scale of systems is beyond the effective Bohr radius{\bf \ }%
\cite{song}.

We assume the wave function of the exciton in the form 
\begin{equation}
\Psi =\sum\limits_{\lambda ,\lambda ^{\prime }}A_{\lambda ,\lambda ^{\prime
}}\psi _\lambda ^{cm}\left( {\bf \vec{R}}\right) \psi _{\lambda ^{\prime
}}^{rel}\left( {\bf \vec{r}}\right) ,  \label{wvfForm}
\end{equation}
where $\psi _\lambda ^{cm}\left( {\bf \vec{R}}\right) $ and $\psi _{\lambda
^{\prime }}^{rel}\left( {\bf \vec{r}}\right) $ are the respective wave
functions of the center-of-mass and the relative Hamiltonians $H_{cm}$ and $%
H_{rel}$, which can be solved by the series expansion method introduced by
Zhu {\em et al}. \cite{zhu90,zhu97}. $\lambda =\left\{ n_{cm},m_{cm}\right\} 
$ and $\lambda ^{\prime }=\left\{ n_{rel},m_{rel}\right\} $ represent the
quantum number pairs of the radial quantum number $n$ and orbital angular
momentum quantum number $m$. Due to the cylindrical symmetry of the problem,
the total orbital angular momentum $L=m_{cm}+m_{rel}$ is a good quantum
number of the exciton wave functions.

To obtain the coefficients $A_{\lambda ,\lambda ^{\prime }}$, the total
Hamiltonian is diagonalized in the space spanned by the product states $\psi
_\lambda ^{cm}\left( {\bf \vec{R}}\right) \psi _{\lambda ^{\prime
}}^{rel}\left( {\bf \vec{r}}\right) $. First of all, we solve the single
particle problem of center-of-mass and relative Hamiltonians $H_{cm}$ and $%
H_{rel}$, keep several hundreds of the single particle states, and then pick
up the low-lying energy levels to construct several thousands of product
states. Note that our numerical diagonalization scheme is very efficient and
essentially exact in the sense that the accuracy can be improved as required
by increasing the total number of selected low-lying energy levels $f$. For
instance, when the product states $f=1024$ are kept in $L=0$ subspace, the
precision of the ground state energy has a relative convergence of $\sim
10^{-6}$.

Once the coefficients $A_{\lambda ,\lambda ^{\prime }}$ are obtained, one
can calculate directly the measurable properties, such as the linear optical
susceptibility of the nano-rings, whose imaginary part is related to the
absorption intensity measured by the near-infrared transmission. In theory,
the linear optical susceptibility is proportional to the dipole matrix
elements between one electron-hole pair $j$ state and the vacuum state,
which in turn is proportional to the oscillator strengths $F_j$. In the
dipole approximation, it is given by \cite{que,song,bryant} 
\begin{eqnarray}
F_j &=&\left| \int \int d{\bf \vec{R}}d{\bf \vec{r}}\Psi \left( {\bf \vec{R},%
\vec{r}}\right) \delta \left( {\bf \vec{r}}\right) \right| ^2  \nonumber \\
&=&\left| \sum\limits_{\lambda ,\lambda ^{\prime }}A_{\lambda ,\lambda
^{\prime }}\psi _{\lambda ^{\prime }}^{rel}\left( {\bf 0}\right) \int d{\bf 
\vec{R}}\psi _\lambda ^{cm}\left( {\bf \vec{R}}\right) \right| ^2,
\label{oscStrength}
\end{eqnarray}
where the factor $\psi _{\lambda ^{\prime }}^{rel}\left( {\bf 0}\right) $
and the integral over ${\bf \vec{R}}$ ensure that only the excitons with $%
L=0 $ (or more precisely $m_{cm}=m_{rel}=0$) are created by absorbing
photons. Therefore, the frequency dependence of the linear optical
susceptibility $\chi \left( \omega \right) $ can be expressed as \cite
{que,song,bryant} 
\begin{equation}
\chi \left( \omega \right) \propto \sum_jF_j\cdot (\hbar \omega
-E_g-E_j-i\Gamma )^{-1},  \label{kapa}
\end{equation}
where $E_g$ and $E_j$ are the respective semiconducting band gap of InAs and
the bound state energy levels of the exciton, and $\Gamma $ has been
introduced as a phenomenological broadening parameter.

In what follows we limit ourselves in the subspace $L=0$. As an interesting
example, we choose the parameters $m_e^{*}=0.067m_e$ \cite{lorke00,emperador}%
, the effective mass of the light hole $m_{lh}^{*}=0.099m_e$, the effective
mass of the heavy hole $m_{hh}^{*}=0.335m_e$, and the appropriate parameter
to InGaAs $\varepsilon _r=12.4$. Moreover, the electron and hole are
considered to be confined in the a confinement potential with the same
strength, {\em i.e.}, $m_e^{*}\omega _e^2=m_h^{*}\omega _h^2$. If the
characteristic energy and length scales are the effective Rydberg radius $%
R_y^{*}=\frac{\mu e^4}{2\hbar ^2\left( 4\pi \varepsilon _0\varepsilon
_r\right) ^2}$ and the effective Bohr radius $a_B^{*}=\frac{4\pi \varepsilon
_0\varepsilon _r\hbar ^2}{\mu e^2}$ for the heavy hole exciton, we find $%
R_y^{*}=5.0$ {\rm meV} and $a_B^{*}=11.8$ {\rm nm}. In order to simulate the
InAs nano-rings with $\Delta E_m\sim 5$ {\rm meV} and $\Delta E_n\sim 20-25$ 
{\rm meV} in the experimental measurements \cite{pettersson}, $R_0$ and $%
\hbar \omega _e$ are chosen to be $20$ {\rm nm} and $14$ {\rm meV}
respectively ($\hbar \omega _e=14${\bf \ }{\rm meV} corresponds to a ring
width $W=10$\ {\rm nm}). $\Delta E_m$ is the energy level spacing between
the single electron states with different orbital angular momentum $m$ and
the same radial quantum number $n$, while $\Delta E_n$ corresponds to the
energy spacing with different radial quantum number $n$ and the same angular
momentum $m$.

In Fig.2a and Fig.2b, the {imaginary part of linear optical susceptibility }%
with two different ring radii: $R_0=20$ {\rm nm }and{\rm \ }$5${\rm \ nm} is
shown{\rm , }where a broadening parameter $\Gamma =2.5$ {\rm meV} is used.
The solid and dashed lines correspond to the heavy hole and light hole
excitons. Those curves represent all the possible transitions of excitonic
states which can be measured by photoluminescence excitation measurements
(PLE){\bf .} Comparing these two figures, one finds the following
differences: i). For a large ring radius, only a few low-lying states
transitions appear in the low photon energy regime. As photon energy
increases, the intensity of transitions damps rapidly. ii). For a small ring
radius, the low-lying states transitions is nearly equally distributed in
the whole photon energy regime both for the heavy-hole and light-hole
excitons. iii). The intensity magnitude of the ground state transition in
Fig.2b is almost one order smaller than that in Fig.2a. All these features
amazingly coincide with the experimental observations between nano-rings and
quantum dots structures \cite{pettersson}. Indeed, for the nano-ring with a
smaller radius $5${\rm \ nm}, its width is comparable to the radius, $W\sim
2R_0$. The description of the small radius nano-ring should be equivalent to
that of a quantum dot. The {imaginary part of linear optical susceptibility}
for a cylindrical quantum dot in a parabolic potential $U\left( {\bf \vec{r}}%
\right) =\frac 12m^{*}\omega _0^2r^2$ ($\hbar \omega _0$ is taken to be $28$ 
{\rm meV}) is also displayed in Fig.2c, whose features are very similar to
that of Fig.2b.

The distinct differences between nano-rings and quantum dots mainly
originate from their geometries. For quantum dots, nearly periodic
distributed low-energy transitions (the higher peaks in Figs. 2b or 2c) are
reflections of electron-hole excitations involving the exciton ground state
and various center-of-mass replicas without altering the ground state of the
relative coordinate, while the smaller amplitude peaks of the photon energy
at $1.341,1.362,1.368$\ eV in Fig. 2b correspond to {\it internal} excited
states of the exciton (its relative coordinate){\bf \ }\cite{note}. However,
for quasi-one-dimensional nano-rings, this situation is totally changed. The
center-of-mass degrees of freedom are greatly suppressed by the anisotropic
ring-like confinement and the relative motion becomes dominant, giving rise
to the destruction of the regular patterns observed in quantum dots.{\bf \ }

To fully understand the effects of ring radius $R_0$ on the optical
properties, the transition peak's positions or the low lying energy levels
of excitons and the corresponding oscillator strengths are plotted as a
function of the ring radius in Fig.3a and Fig.3b, respectively. Each line is
delineated by the transition sequence indicated in Fig.2b. The most striking
feature is that the oscillator strengths $F_j$ are approximately
proportional to the ring radius. Amazingly, for $R_0=20$ {\rm nm, }the
oscillator strengths of the ground state and first excited transitions are
nearly $30$ and $10$, in good agreement with the experimental values $F_j$ $%
=31$ and $18$ \cite{pettersson}. Moreover, in Fig.3a and Fig.3b, the peak's
positions show a few changes of the low lying transitions when $R_0\geqslant
10$ {\rm nm}.

Let us now compare the theoretical results with the near-infrared
transmission data of H. Pettersson {\em et al. }\cite{pettersson}. In
experiment, one cannot distinguish the contribution from the heavy hole or
the light hole excitons. To reproduce the experimental result for an InAs
nano-ring, in Fig.4b we plot the combination of {imaginary part of linear
optical susceptibility of the heavy-hole and the light-hole excitons with
the ring radius }$R_0=20$ {\rm nm}. Actual self-assembled nano-rings display
a size distribution which greatly affects their physical features because
each nano-ring is very small. To account for the experimental data, here a
relatively large level broadening of $\Gamma =5$ {\rm meV} has been used. By
fitting the rightmost peak's position in Fig. 4a, we take the band-gap $E_g$
to be $1.325$ {\rm eV}. As shown in Fig.4, the overall trend of the
experiment result is qualitatively reproduced by the model calculation, {\em %
i.e.}, both of them exhibit three transition peaks, and the peak's positions
and relative amplitudes agree with each other {\it qualitatively}{\bf .} The
main discrepancy between the present theory and experiment occurs at high
photon energy $1.4$ {\rm eV}, where the experimental curve displays a steep
increase, while the theory predicts no variation. This discrepancy can be
attributed to the increasing contribution from the wetting layer \cite
{private}. The small split of the rightmost peak is somewhat of a biasing
artifact due to the unknown ratio of the contributions of the heavy hole and
light hole excitons. However, considering the simplicity of our
effective-mass model and the use of only one fitting parameter $E_g$, the
overall fit is still quite surprising. Actually, more convincing fits should
involve a suitable set of effective mass parameters, a careful adjusted
confinenent strength, the realistic size distribution, the ratio of heavy
hole and light hole excitons, and others unknown factors in experiments.

Recently, Warburton {\em et al}. presented a beautiful experiment of optical
emission in a single charge-tunable nano-ring \cite{warburton}. They studied
the role of multiply-charged exciton complexes with no applied magnetic
field and found a shell structure in energy similar to that of quantum dots 
\cite{bayer,hawprl00}. Therefore, encouraged by the rapid developed
nano-techniques for detection, i.e., the achievement of extremely narrow and
temperature insensitive luminescence lines from a single InAs nano-ring in
GaAs, we hope that our explanation of the distinct optical properties of
quantum dots and nano-rings can be further confronted in a more precise
experiment in the future{\bf .}

In conclusion, we have studied the low lying exciton states and their
optical properties in a self-assembled semiconducting InAs nano-ring. By
varying the radius $R_0$, both the conventional quantum dots and nano-rings
have been considered within the same framework of our theory. The distinct
optical properties of quantum dots and nano-rings observed in recent
experiments are well explained qualitatively.

{\bf Acknowledgments}

One of the authors (H. Hu) would like to thank Professor A. Lorke for
sending a copy of the paper (Ref. \cite{pettersson}) and Dr. H. Pettersson
for fruitful discussions. We also acknowledge for the financial support from
NSF-China (Grant No.19974019) and China's ''973'' program.

\smallskip

\begin{center}
{\bf Figures Captions}
\end{center}

Fig.1. The confinement potential $U\left( {\bf \vec{r}}\right) =\frac{%
m_e^{*}\omega _e^2}{2R_0^2}\left( {\bf \vec{r}}{}^2-R_0^2\right) ^2$ with
different ring radii with $m_e^{*}=0.067m_e$ and $\hbar \omega _e=14$ {\rm %
meV.}\newline

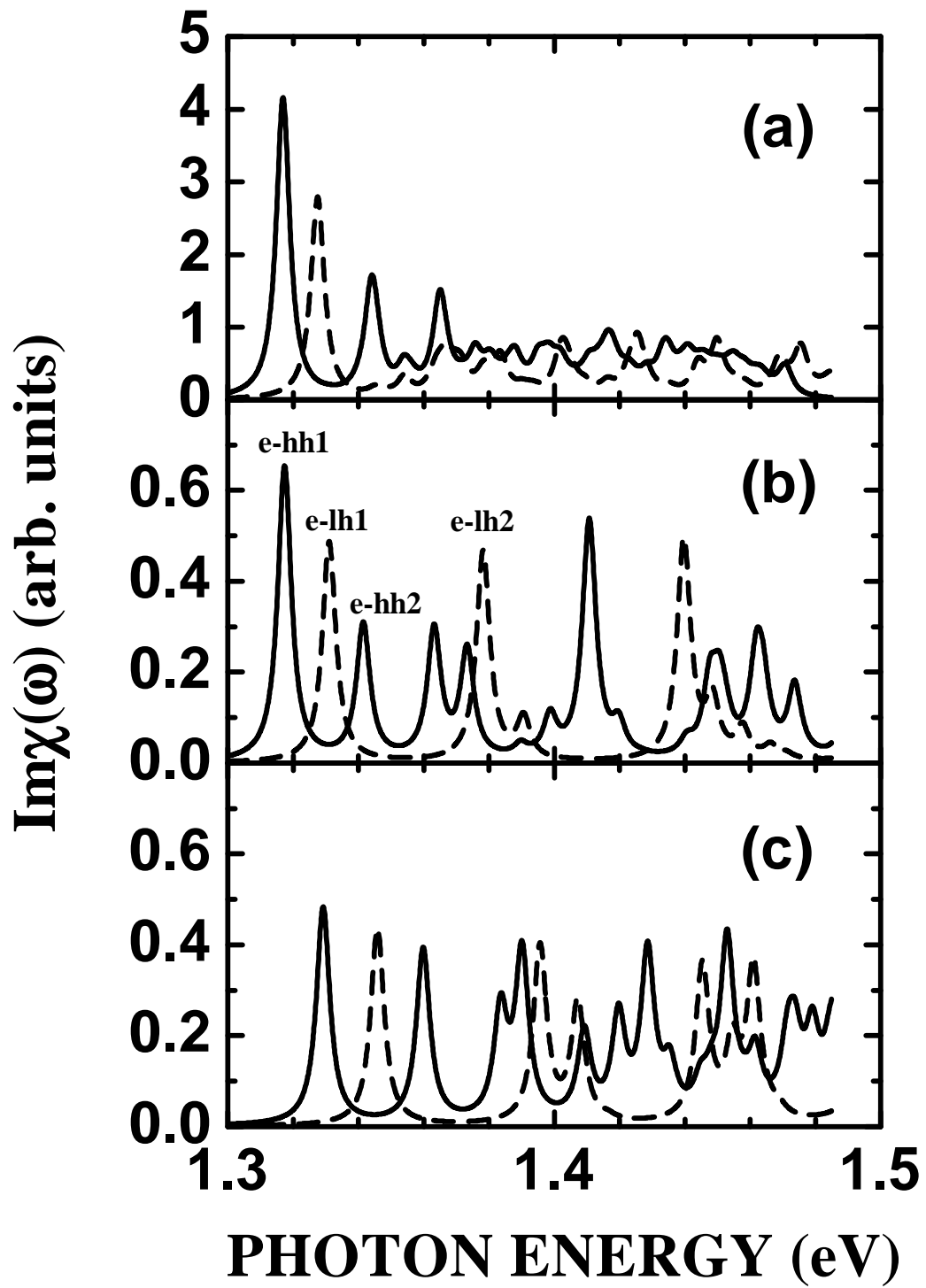
Fig.2. Imaginary part of linear optical susceptibility for the nano-ring as
a function of photon energy with two different ring radii $R_0=20$ {\rm nm}
(a) and $R_0=5$ {\rm nm }(b). For comparison, the results for a parabolic
quantum dot are also displayed (c). The solid and dashed lines correspond to
the heavy-hole and light-hole excitons, respectively. The semiconducting
band gap $E_g$ is $1.325$ {\rm eV}.\newline

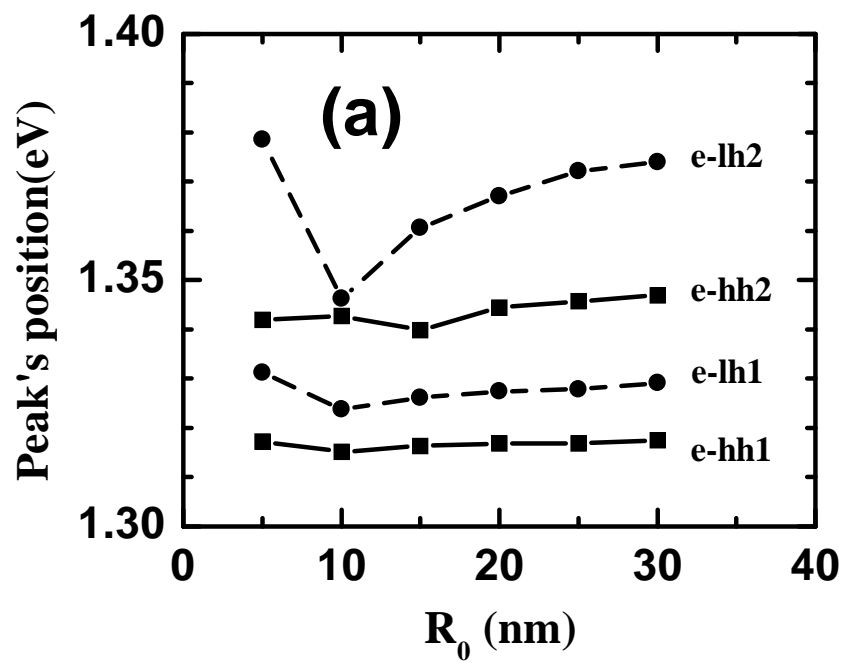
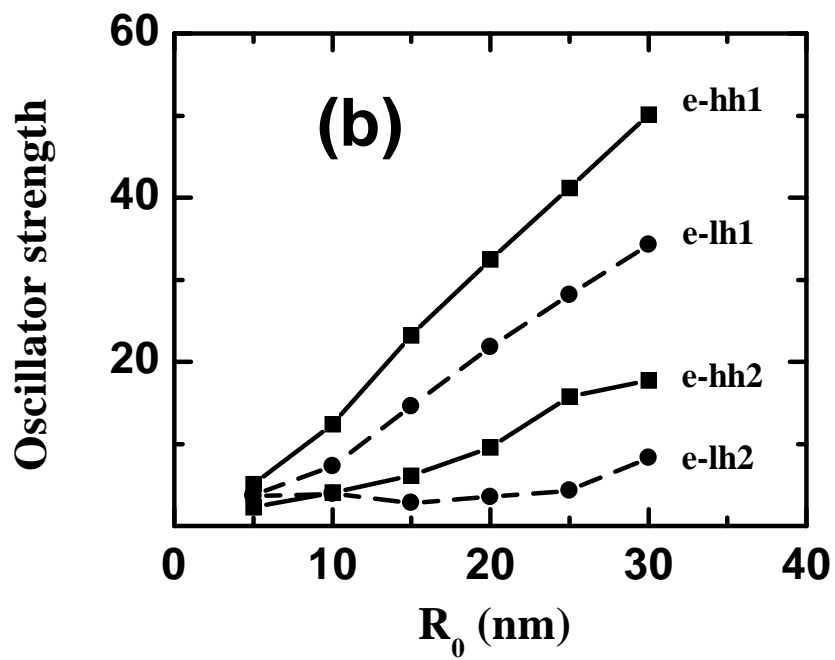
Fig.3. The peak's positions of the linear optical susceptibility (a) and
oscillator strengths (b) {for the }nano-ring as a function of the ring
radius $R_0$. The square and solid circle symbols denote the heavy-hole and
light-hole excitons, respectively. Each line is labeled by the sequence
indicated in Fig.2(b).\newline

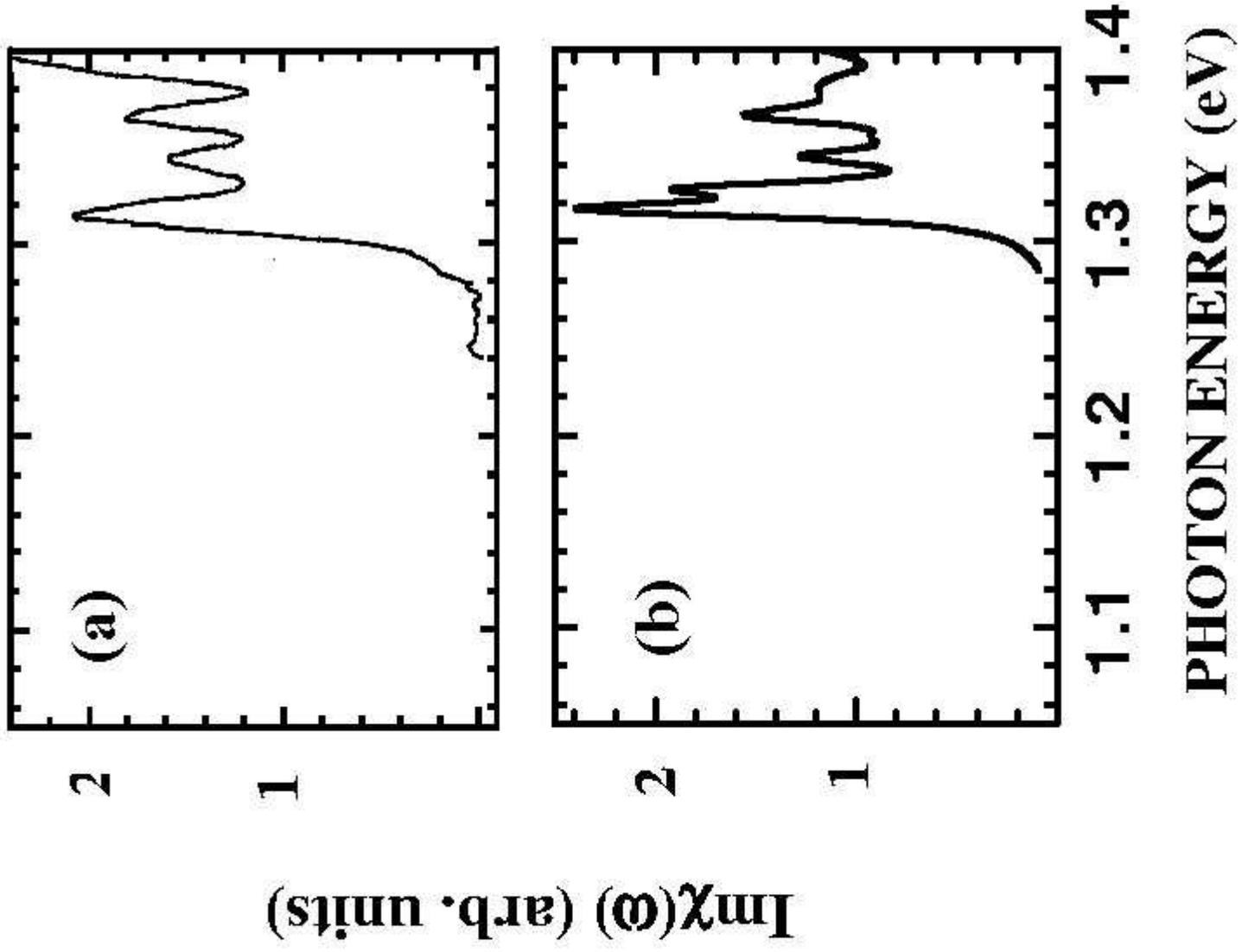
Fig.4. The spectrum of the linear optical susceptibility of the nano-ring as
a function of photon energy. (a) is the experimental result and (b) is our
theoretical one.

\end{document}